\documentclass[preprint,3p]{elsarticle}

\usepackage{amssymb}
\biboptions{sort&compress}
\usepackage{amsmath}
\usepackage[colorlinks,linkcolor=blue,anchorcolor=blue,citecolor=blue]{hyperref}
\usepackage{booktabs, array, mathptmx, float, tabularx, booktabs, lipsum, multirow}
\usepackage{dcolumn}
\usepackage{subcaption}
\usepackage{caption}
\usepackage{bm}

\usepackage[font=small,labelfont=bf,labelsep=none]{caption}  

\usepackage{lineno}
\usepackage{cuted}
\UseRawInputEncoding

\journal{Nuclear Inst. and Methods in Physics Research, A}

\begin{document}
	
	\begin{frontmatter}
		
		
		
		\title{Numerical investigations of heavy ion driven plasma wakefield acceleration} 
		
		
		\author[1,2]{Jiangdong Li}
		\author[1,2]{Jiancheng Yang\corref{cor1}}
		\cortext[cor1]{Corresponding author at: Institute of Modern Physics, Chinese Academy of Sciences, Lanzhou 730000, China.}
		\ead{yangjch@impcas.ac.cn}.
		\author[3,4]{Guoxing Xia}
		\author[1,2]{Jie Liu}
		\author[1]{Wenlong Zhan}
		\author[1,2]{Ruihu Zhu}
		
		\affiliation[1]{
			organization={Institute of Modern Physics, Chinese Academy of Sciences},
			city={Lanzhou},
			postcode={730000},
			country={China}
		}
		\affiliation[2]{
			organization={University of Chinese Academy of Sciences},
			city={Beijing},
			postcode={100049},
			country={China}
		}
		\affiliation[3]{
			organization={University of Manchester},
			city={Manchester},
			postcode={M13 9PL},
			country={United Kingdom}
		}
		\affiliation[4]{
			organization={Cockcroft Institute},
			city={Cheshire},
			postcode={WA4 4AD},
			country={United Kingdom}
		}
		\begin{abstract}
			Plasma-Based Acceleration (PBA) has emerged as a promising approach to achieve ultra-high gradient particle acceleration. While extensive PBA studies have been conducted using laser, electron, and proton drivers, significant challenges remain in achieving high efficiency, stable acceleration, and scalable energy gain. Meanwhile, due to their higher beam charge density, heavier particle mass and higher kinetic energy, heavy-ion beam drivers represent an interesting direction in PBA research. In this paper, the plasma wakefield acceleration driven by heavy ion beam is studied for the first time, aiming to find the best mechanism for generating high-amplitude wakefields. Using the high intensity, high energy heavy ion beams provided by the High Intensity heavy-ion Accelerator Facility (HIAF), our simulations show that heavy ions can excite stable, high-amplitude plasma wakefields up to 6 GV/m, suitable for electron acceleration. These results show good performance of heavy ion beam drivers and their potential as a viable and promising approach in the field of PBA.
		\end{abstract}
		
		
		
		
	\end{frontmatter}
	
	
	
	\section{\label{section1}Introduction}
	\subsection{\label{sectionA}Plasma based acceleration}
	
	Plasma-based acceleration (PBA) is a frontier technique that can achieve much higher acceleration gradients than conventional radio-frequency (RF) based accelerators. The concept of PBA was first proposed by Tajima and Dawson in 1979 \cite{1}, and since then, many theoretical and experimental studies have been conducted to explore its full potential. 
	
	In PBA, a high-energy driver beam passes through a plasma, loses its energy, and excites a wakefield that can accelerate another particle beam, the "witness beam" behind it. One of the main challenges of PBA is to generate a suitable driver beam that can excite strong and stable wakefield in the plasma. Laser driven wakefield acceleration (LWFA), has been widely explored due to availability of ultrashort, high intensity laser pulses \cite{2,3,4}. In recent experiments, using a peak laser power of 850 TW, electron beam pulse with peak energy up to 7.8 GeV can be generated in a distance of 20 cm \cite{5}. Later, electron bunches can be accelerated up to 10 GeV with high-quality guiding of 500 TW laser pulses over 30 cm at the BELLA Center in Berkeley Lab \cite{Picksley:2024cdd}. In order to accelerate an electron bunch to 1 TeV, these accelerating gradients would have to be maintained over distances of tens of meters, or many acceleration stages would have to be combined. It is a great challenge for relative timing, alignment, and matching between these acceleration stages. In addition, the effective gradient is reduced because of the introduction of long sections, which are used for injections of the driver beam and dumping of the spent beam between the acceleration stages. 
	
	Later, it was recognized that the plasma wakefield could also be excited by the relativistic charged particle beam, such as an electron beam \cite{7,8}. Using an electron beam with sufficient intensity, the wakefields will be excited in plasma along its propagation path \cite{9}. An experiment conducted at SLAC demonstrated a wakefield of 52 GV/m was achieved and sustained for almost 85 cm \cite{10}. Then, another experiment at SLAC injected a trailing electron beam behind the driver beam. The results show that particles in the trailing bunch with charge of 74 pC can be accelerated to 1.6 GeV over 36 cm with a good energy spread of 0.7 \%. The energy-transfer efficiency from the wake to the bunch in this experiment exceeds 30 \% \cite{11}. However, the maximum energy gain for the witness beam in any symmetric bunch distribution is limited by the transformer ratio, which is less than 2 \cite{12}. 
	
	On the other hand, proton-driven plasma wakefield acceleration has also been proposed and successfully demonstrated in AWAKE (Advanced Wakefield Experiment) at CERN \cite{13,14,15,16}. A long and thin proton bunch was used because it undergoes a particle-plasma interaction called self-modulation instability. This interaction splits the long proton bunch longitudinally into a series of high-density microbunches, which then excite resonantly high amplitude wakefields \cite{17,18,19}. AWAKE successfully accelerated electrons up to 2 GeV in 10 m-long plasma, and proved this technique has the potential to accelerate electrons to the TeV scale in a single acceleration stage \cite{20}.
	
	\subsection{\label{sectionB}Heavy ion beams as drivers}
	
	Heavy ion beam drivers have not yet been extensively explored for applications in plasma wakefield acceleration (PWFA), but they possess several potential benefits. The high kinetic energy of heavy ion beams implies that a significant amount of energy can be transferred to the witness beam. With higher beam charge density, heavy ion beams can deposit more energy in the plasma and excite wakefields with higher amplitudes. Moreover, the heavier particle mass of heavy ion beams means that they can sustain stable wakefields over longer distances in the plasma.
	
	High Intensity heavy-ion Accelerator Facility (HIAF) in China features the highest pulse current intensity of heavy ion beams \cite{21}. The kinetic energy can reach MJ level. Furthermore, HIAF is going to be commissioned in 2025. Within a year or two, HIAF will be able to provide high energy, high intensity heavy ion beams and will be ready for a heavy ion driven plasma wakefield acceleration experiment in the next few years. Therefore, it is very timely to use heavy ion beams from HIAF to simulate plasma wakefield acceleration. By employing heavy ion beams, it is possible to achieve more energy gain in the witness beam within a single stage of plasma acceleration, which is desirable for high energy colliders. 
	
	This paper is structured as follows: Section 2 analyzes the excitation of plasma wakefields driven by heavy ion beams. Section 3 details the simulations of stable wakefield formation using various heavy ion drivers from existing and under-construction heavy-ion accelerator facilities in China. Section 4 simulates the electron acceleration under optimized wakefield conditions. Conclusions are summarized in Section 5.
	
	\section{\label{section2}Principle of plasma wakefield excitation}
	
	In order to drive plasma wakefield more efficiently, the RMS transverse size of the driver beam should satisfy 
	
	\begin{equation}\label{eq:satisfy_condition1}
		\frac{c}{\omega_{pe}} \gtrsim \sigma_r
	\end{equation}
	or
	
	\begin{equation}\label{eq:satisfy_condition2}
		k_{pe} \sigma_r \lesssim 1.
	\end{equation}
	where $k_{pe} = 2\pi / \lambda_{pe}$, $\lambda_{pe}$ is the plasma electron wavelength and it is given by $\lambda_{pe} = 2\pi c / \omega_{pe}$. This prevents the plasma return current from flowing within the beam, suppressing the development of the current filamentation instability (CFI) and preserving plasma wakefield development \cite{22}. The plasma electron oscillation frequency $\omega_{pe}$ is related to the plasma electron density $n_{pe}$ as
	
	\begin{equation}\label{eq:plasma_freq}
		\omega_{pe} = \sqrt{\frac{n_{pe} e^2}{\epsilon_0 m_e}}
	\end{equation}
	where $m_e$ is the electron mass and $\epsilon_0$ is the vacuum permittivity. The highest achievable accelerating field is on the order of the wave breaking field $E_{WB}$:
	
	\begin{equation}\label{eq:wave_breaking_field}
		E_{WB} = m_e c \omega_{pe} / {e}
	\end{equation}
	By combining Eq. (\ref{eq:plasma_freq}) and (\ref{eq:wave_breaking_field}), the wave breaking field can be estimated by $E_{WB} \approx 96 \sqrt{n_{pe} \left( \text{cm}^ \text{-3} \right)}$ V/m. For a GV/m accelerating wakefield, the plasma density $n_{pe}$ should exceed $10^ \text{14}  \text{cm}^ \text{-3}$, corresponding to a plasma wavelength $\lambda_{pe}$ of approximately 3 mm, and the optimal beam radius is typically around 0.5 mm. According to \cite{23}, the optimal wake would be obtained for $k_{pe} \sigma_z \approx \sqrt{2}$. Thus, the driver beam length $\sigma_z$ should also be less than 1 mm.
	
	However, the bunch lengths from the current heavy ion facilities are on the order of a few meters, which are too long to efficiently drive a high amplitude plasma wakefield. To facilitate heavy ion driven plasma wakefield acceleration, there are two main strategies to effectively drive the wake: either compress the entire driver beam or modulate the driver beam into microbunches to achieve resonant wakefield excitation. Extensive research \cite{magneticcompression} has demonstrated the significant challenges in achieving effective beam compression, so only the latter is discussed in the following sections.
	
	When a long heavy ion bunch ($\sigma_z \geq \lambda_{pe}$) enters a plasma, its head generates a wakefield that acts on its tail. The periodic regions of focusing and defocusing modulate the beam density at $\lambda_{pe}$, driving a larger plasma density modulation and further leading to an unstable modulation of the whole bunch along the bunch propagation direction. This self-modulation will split a long heavy ion bunch into many microbunches with distance between two adjacent microbunches being about $ \lambda_{pe}$ to resonantly drive the plasma wake. This process is called self-modulation instability (SMI) \cite{17}. The self-modulation instability is mainly due to the influence of the transverse wakefields acting on the bunch itself. SMI is different from the electrostatic two-stream instability, which arises due to relative streaming between the bunch and the background plasma in longitudinal.
	
	In the linear wake regime, assuming that $r_0$ is the beam equilibrium radius, the beam radius $r_b = r_0 +r_1$ with $|r_1/r_0| \ll 1$. By letting $r_{b0} = r_0$, $r_1$ is the slowly varying amplitude of the beam radius perturbation. The evolution of $r_1$ can be expressed as \cite{25}
	
	\begin{equation}\label{eq:beam_radius_perturbation}
		r_1 = \delta r \frac{3^{\text{1/4}}}{\left( 8 \pi\right)^{\text{1/2}}} N^{\text{-1/2}} e^N \cos(\frac{\pi}{12} - k_{pe} \xi - \frac{N}{\sqrt{3}})
	\end{equation}
	
	\begin{equation}\label{eq:e_foldings}
		N = \frac{3^{\text{3/2}}}{4} \left( \nu \frac{n_{\text{b0}} m_e}{n_0 M_b \gamma} k^3_{pe} |\xi| z^2 \right)^{\text{1/3}}
	\end{equation}
	
	The self-modulation growth rate is given by Eq. (\ref{eq:e_foldings}) with $|\xi| = L_b$, $L_b$ is the bunch length. Hence, for fixed $r_{b0}$, the growth rate scales as $N \propto \left( n_{b0} L_b / M_b \right)^{\text{1/3}} \propto \left( Q_b / M_b \right)^{\text{1/3}}$, where $Q_b$ is the beam charge per bunch and $M_b$ is the beam particle mass. Therefore, although typically exhibiting a high charge state, heavy-ion beams are disadvantageous for the development of self-modulation instability due to their heavier particle mass. However, when the energy of the witness beam exceeds the GeV range, the self-modulation distance becomes negligible compared to the acceleration distance of the witness beam, which means the issue of self-modulation growth rate is not a critical factor.
	
	Several analytical models and mathematical expressions for the SMI growth rates are developed (such as Eq. (\ref{eq:e_foldings})), numerical simulations show that the microbunches generated by the SMI can resonantly drive a high amplitude wakefield ($\sim$ GV/m). However, the propagation of the driver beam in the plasma will also be affected by other instabilities, which may significantly disrupt or even destroy the wakefields. Fortunately, these instabilities can be mitigated through seeded self-modulation (SSM) \cite{26,27,28,29}. In this approach, self-modulation is not initiated by random shot noise or minor bunch irregularities. Instead, it can be achieved by using a preceding bunch to generate the initial transverse wakefield or by employing a charge distribution with a sharp leading edge. This reduces the propagation distance required to develop high-amplitude wakefields and further suppresses other instabilities.
	
	\section{\label{section3}Simulations of stable wakefield formation}
	\subsection{\label{sectionA}Quasi-static LCODE PIC code}
	
	In this section, a particle-in-cell (PIC) code named "LCODE" will be used for simulations of particle beam-driven plasma wakefield acceleration \cite{30}. LCODE operates in 2-dimensional geometry (2D3V), supporting both planar and axisymmetric geometries. The code uses a light-speed co-moving simulation window and quasi-static approximation for plasma response calculations. Particle beams are represented by fully relativistic macroparticles. The plasma is simulated either through macroparticle kinetics or a fluid dynamics approach. The kinetic solver enables modeling of transversely inhomogeneous plasmas, hot plasmas, non-neutral plasmas, and accounts for mobile ion dynamics. Comprehensive diagnostic tools are integrated into the code, enabling graphical visualization of simulation results.
	
	The quasi-static approximation is illustrated by treating the beam as "rigid" when calculating the plasma response. In this framework, electromagnetic fields are expressed as functions of the transverse coordinate $r$ and the co-moving coordinate $\xi = z - ct$, and are computed layer-by-layer starting from the beam head. These fields are subsequently used to update beam particle trajectories. For highly relativistic beams, the time step $\Delta t$ for tracking beam particles can be significantly enlarged, accelerating simulations by several orders of magnitude. Crucially, the quasi-static approximation remains valid only if the characteristic timescale of beam evolution is much longer than the period of plasma wave.
	
	The code allows many initial distributions of beam particles over the simulation window, and this paper uses the initial beam density of the form \cite{31}
	
	\begin{equation}\label{eq:beam_density_cos}
		n_b = \frac{n_{b0}}{2} e^{{- r^2} / {2 {\sigma_r}^2}} \left[ 1 - \cos(\frac{2\pi \xi}{L}) \right] \quad \text{,} \quad -L < \xi < 0.
	\end{equation}
	
	The cosine distribution over $\xi$ is more convenient than the Gaussian one because it smoothly vanishes outside an interval of a finite length. For $L = 2 \sqrt{2\pi} \sigma_z$, the distribution Eq. (\ref{eq:beam_density_cos}) is close to the Gaussian one (Eq. (\ref{eq:beam_density_gaussian})) and contains the same number of particles.
	
	\begin{equation}\label{eq:beam_density_gaussian}
		n_b = n_{b0} exp\left(- \frac{r^2}{2 {\sigma_r}^2} - \frac{{\left( \xi + L / 2\right)}^2}{2 {\sigma_z}^2} \right)
	\end{equation}
	
	\begin{figure}[htbp]
		\centering
		\includegraphics[width=8.6cm]{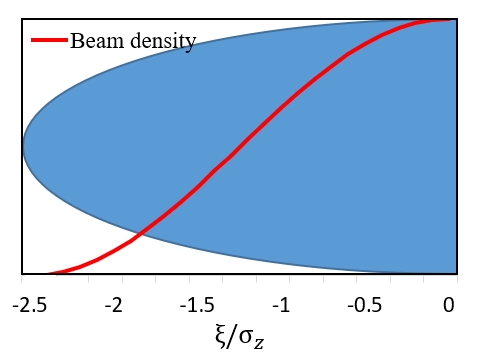}
		\caption{\label{fig:Fig_1}(Color) Schematic of the beam distribution in the simulation window.}
	\end{figure}
	
	\subsection{\label{sectionB}Simulations of stable wakefield formation using different drivers}
	
	Figure \ref{fig:Fig_1} shows the schematic of the beam distribution in the simulation window. In this section, different heavy ion beam drivers were investigated to identify a good regime to achieve a high amplitude plasma wakefield for heavy ion driven plasma wakefield acceleration. To ensure the heavy ion driver highly rigid regardless of the beam radius, the emittance is set to zero.
	
	\subsubsection{Drivers in HIRFL-CSR}
	
	Presently, there are only two facilities in China that have the ability to generate high-energy heavy-ion beams: the existing HIRFL-CSR (Heavy Ion Research Facility in Lanzhou - Cooling Storage Ring) and the forthcoming HIAF (High Intensity heavy-ion Accelerator Facility). Consequently, the drivers in HIRFL-CSR will be discussed first. HIRFL-CSR, is the post-acceleration system of HIRFL \cite{32}. It consists of a main ring (CSRm) and an experimental ring (CSRe). The suggested parameters of drivers are given in Table \ref{tab:drivers}, and the parameters within the parentheses represent the parallel parameters.
	
	\begin{table}[htbp]
		\centering
		\renewcommand{\arraystretch}{1.5}
		\caption{\label{tab:drivers}
			The parameters of drivers in HIRFL-CSR and HIAF.}
		\begin{tabular}{l|c|c|c|c}
			\toprule
			Beam parameters &CSR &\multicolumn{3}{c}{HIAF} \\
			\hline
			Driver Beam& \(^{12}\text{C}^{6+}\) & \(^{12}\text{C}^{6+}\) & \(^{209}\text{Bi}^{83+}\)& p\\
			\hline
			Energy(GeV/u)& 1 & 4.25 & 9.58 & 9.3 \\
			\hline
			Particle Per Bunch& $1 \times 10^{11}$ & $1 \times 10^{12}$ & $1 \times 10^{12}$ & $6 \times 10^{12}$\\
			\hline
			RMS Bunch Radius(mm)& \multicolumn{4}{c}{1(0.1)}\\
			\hline
			RMS Bunch Length(m)& 0.5 & 0.5 & 5 & 0.5\\
			\hline
			Plasma Density($\text{cm}^{-3}$)& \multicolumn{4}{c}{$2.8 \times 10^{13}(2.8 \times 10^{15})$}\\
			\hline
			Wave-breaking Field(GV/m)& \multicolumn{4}{c}{0.5(5.1)}\\
			\hline
			\multicolumn{5}{c}{Simulation parameters in LCODE}\\
			\hline
			Grid step& 0.1 & 0.02 & 0.1 & 0.1\\
			\hline
			Simulation window size(m)& 5(1) & 2(0.2) & 5(1) & 5(1)\\
			\hline
			Particles in layer & \multicolumn{4}{c}{160}\\
			\bottomrule
		\end{tabular}
	\end{table}

\begin{figure}[htbp]
		\centering
		\includegraphics[width=8.6cm]{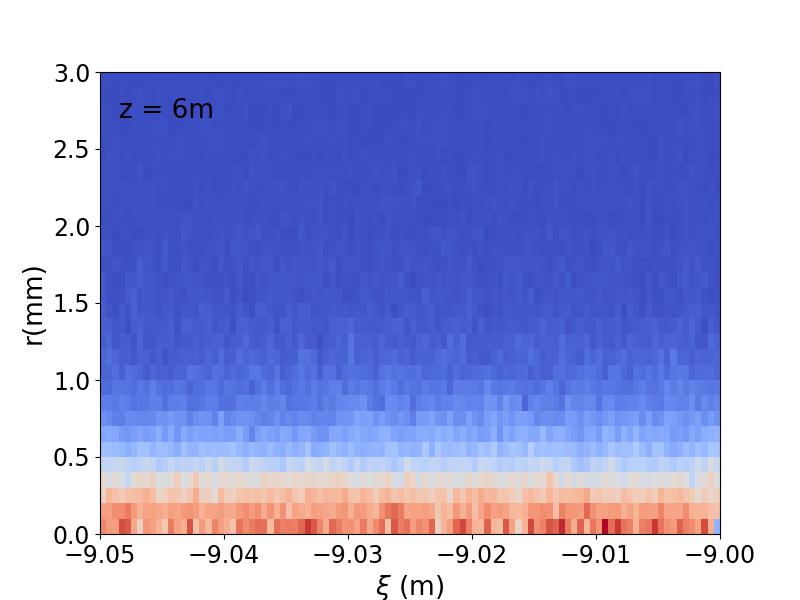}
		\caption{\label{fig:Fig_2}(Color) \(^{12}\text{C}^{6+}\) beam profile (RMS radius = 1 mm) at $z = 6m$ in HIRFL-CSR, in co-moving coordinates $\xi$.}
	\end{figure}

	To find out a good regime for achieving a high amplitude plasma wakefield, the same parameters except the RMS beam radius are used in this simulation. For the carbon beam with an RMS beam radius $=$ 1 mm, no self-modulation instability phenomenon is seen at a distance of 6 m, as shown in Fig. \ref{fig:Fig_2}.
	
	\begin{figure}[htbp]
		\centering
		\subfloat[]{\includegraphics[width=8.6cm]{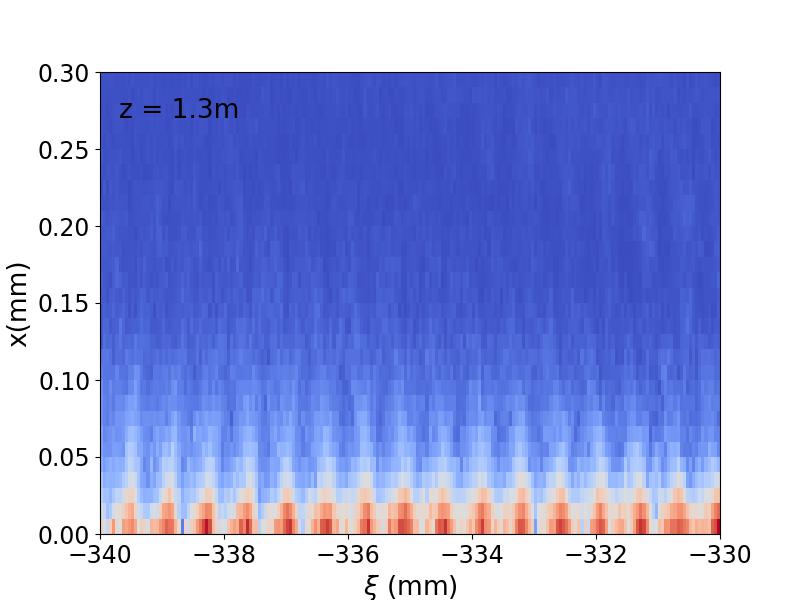}\label{fig:Fig_3a}}
		\subfloat[]{\includegraphics[width=8.6cm]{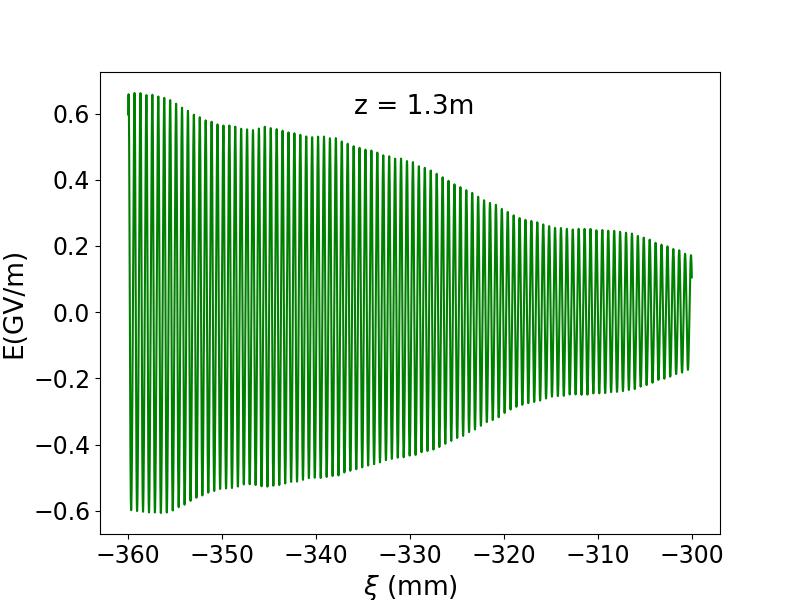}\label{fig:Fig_3b}}
		\caption{(Color) Simulation results of \(^{12}\text{C}^{6+}\) beam (RMS radius = 0.1 mm) in HIRFL-CSR: (a) beam profile and (b) wakefield amplitude with co-moving coordinates $\xi$.}
		\label{fig:Fig_3}
	\end{figure}
	
	For the carbon beam with an RMS beam radius $=$ 0.1 mm, the head of the beam undergoes self-modulation instability within a distance of 1.3 m, resulting in a series of micro-bunch structures and exciting a wakefield with a peak amplitude of 600 MV/m, as shown in Fig. \ref{fig:Fig_3}.
	
	\subsubsection{Drivers in HIAF}
	
\begin{figure}[h]
		\centering
		\includegraphics[width=8.6cm]{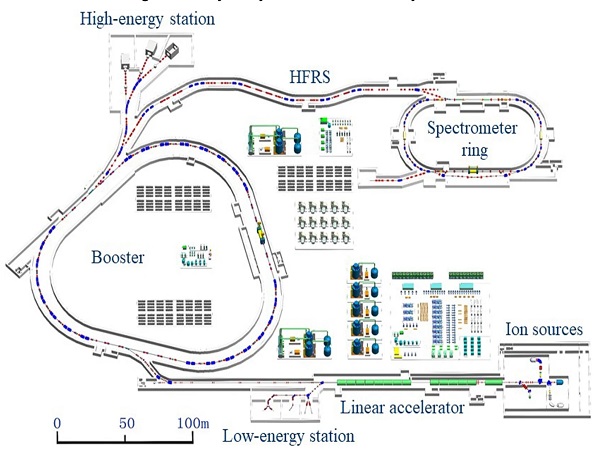}
		\caption{\label{fig:Fig_4}(Color) Layout of HIAF \cite{21}.}
	\end{figure}

	HIAF is a large-scale heavy-ion research facility. As shown in Fig. \ref{fig:Fig_4}, it is composed of the ion sources, the superconducting ion linear accelerator (hereafter abbreviated as linear accelerator), the high-energy synchrotron booster (booster), the high-energy fragment separator (HFRS), the experimental spectrometer ring (spectrometer ring) and the experimental setups. The very intense various stable and radioactive beams with energies from MeV/u to GeV/u provided by HIAF, will give us the best opportunity to explore the potential of heavy ion driven plasma wakefield acceleration. The parameters of the carbon beam in HIAF are summarized in Table \ref{tab:drivers}.
	
\begin{figure*}
		\centering
		\begin{minipage}{0.49\textwidth}
			\includegraphics[width=\linewidth]{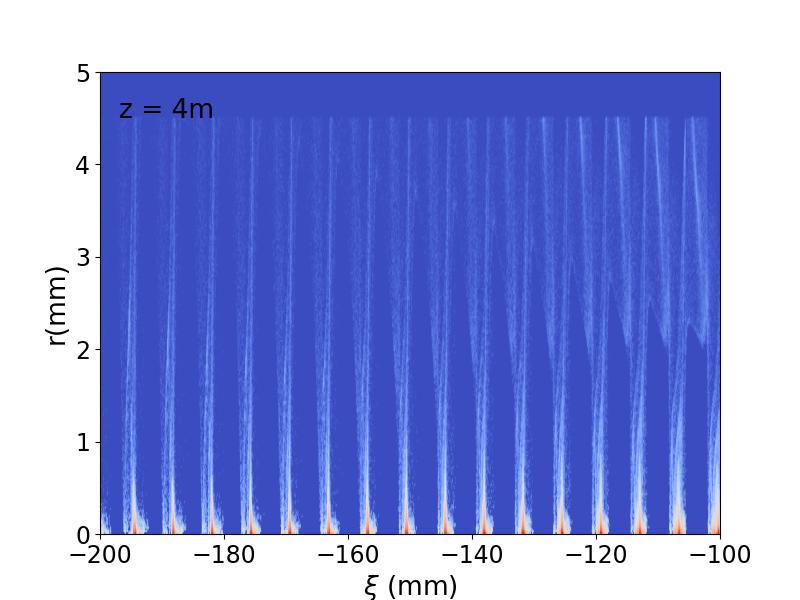}
			\subcaption{}
			\label{fig:Fig_5(a)}
		\end{minipage}
		\hfill
		\begin{minipage}{0.49\textwidth}
			\includegraphics[width=\linewidth]{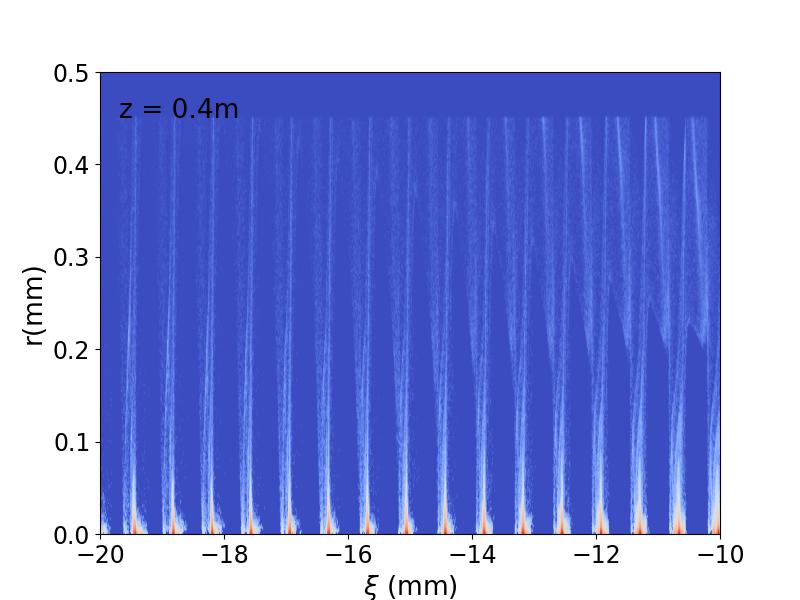}
			\subcaption{}
			\label{fig:Fig_5(b)}
		\end{minipage}
		\vspace{0.01cm}
		\begin{minipage}{0.49\textwidth}
			\includegraphics[width=\linewidth]{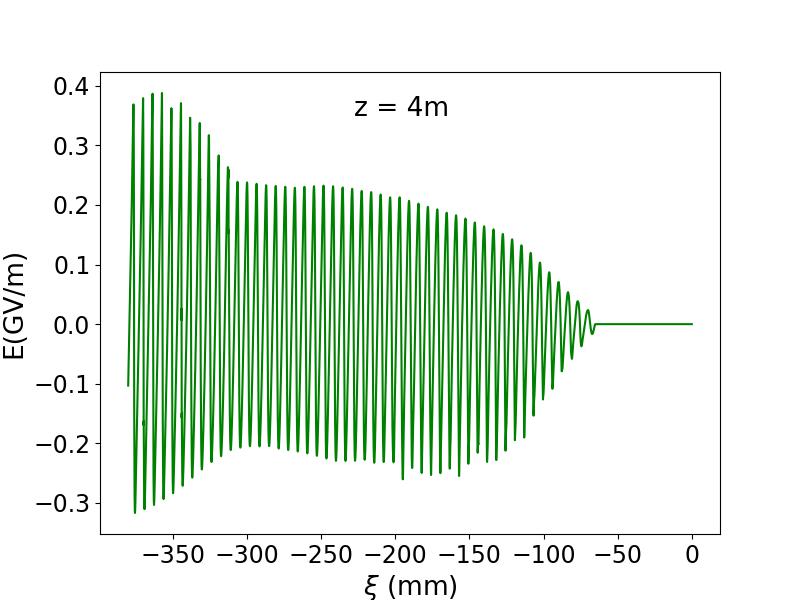}
			\subcaption{}
			\label{fig:Fig_5(c)}
		\end{minipage}
		\hfill
		\begin{minipage}{0.49\textwidth}
			\includegraphics[width=\linewidth]{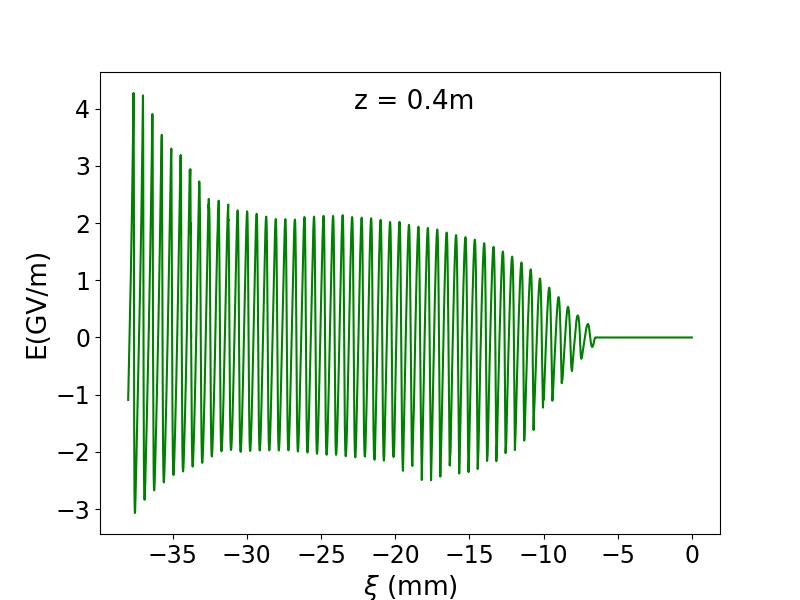}
			\subcaption{}
			\label{fig:Fig_5(d)}
		\end{minipage}
		\caption{(Color) Beam profile (a,b) and the wakefield amplitude (c,d) with co-moving coordinates $\xi$ for \(^{12}\text{C}^{6+}\) beam with an RMS beam radius $=$ 1 mm (left) and \(^{12}\text{C}^{6+}\) beam with an RMS beam radius $=$ 0.1 mm (right).}
		\label{fig:Fig_5}
	\end{figure*}

	Compared with the carbon beam in HIRFL-CSR, the carbon beam in HIAF will have a higher energy and charge intensity, which is better for acceleration and development of self-modulation instability. For the carbon beam with an RMS beam radius $=$ 1 mm, Fig. \ref{fig:Fig_5(a)} and \ref{fig:Fig_5(c)} show that a series of microbunches are formed in the beam head, and there is a wakefield with a maximum amplitude of 400 MV/m after a distance of 4 m. In contrast, the carbon beam with an RMS beam radius $=$ 0.1 mm will generate a stronger wakefield (peak amplitude: 4 GV/m), as seen in Fig. \ref{fig:Fig_5(b)} and \ref{fig:Fig_5(d)}.
	
	Next, the highest energy and highest intensity about heavy ion beam provided by HIAF, specifically \(^{209}\text{Bi}^{83+}\), is used to simulate self-modulation instability. The Bismuth beam will carry an energy of 0.32 MJ and the parameters are summarized in Table \ref{tab:drivers}.
	
\begin{figure*}
		\centering
		\begin{minipage}{0.49\textwidth}
			\includegraphics[width=\linewidth]{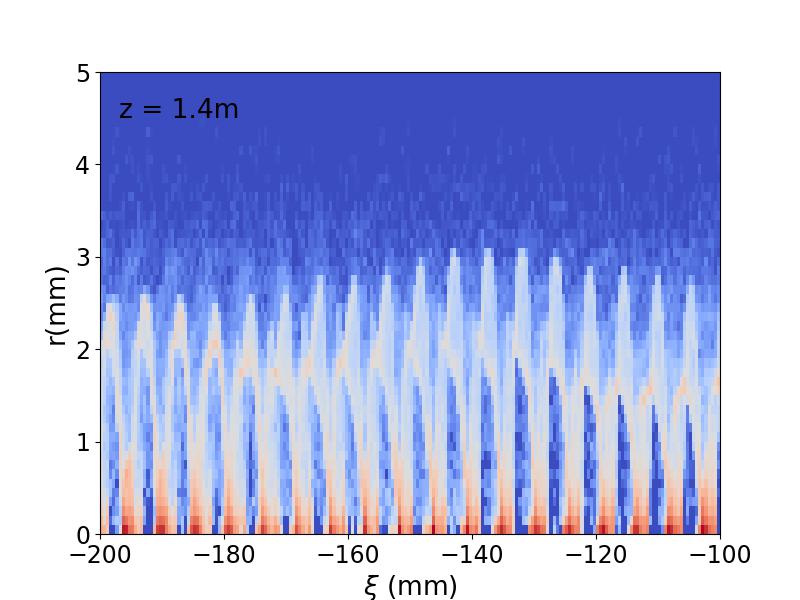}
			\subcaption{}
			\label{fig:Fig_6(a)}
		\end{minipage}
		\hfill
		\begin{minipage}{0.49\textwidth}
			\includegraphics[width=\linewidth]{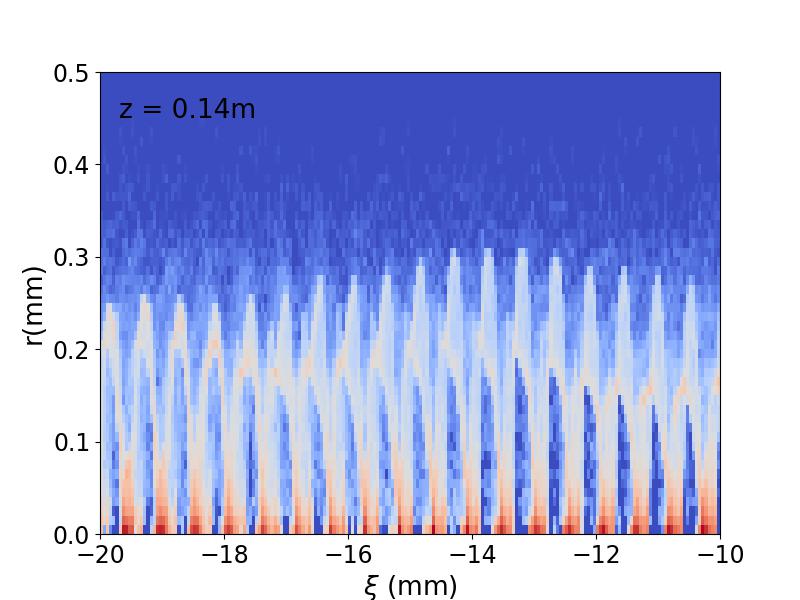}
			\subcaption{}
			\label{fig:Fig_6(b)}
		\end{minipage}
		\vspace{0.01cm}
		\begin{minipage}{0.49\textwidth}
			\includegraphics[width=\linewidth]{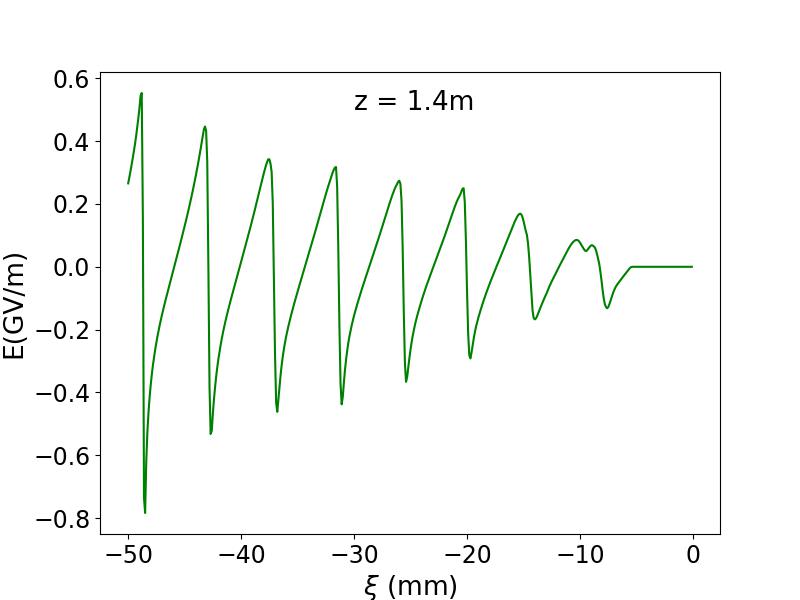}
			\subcaption{}
			\label{fig:Fig_6(c)}
		\end{minipage}
		\hfill
		\begin{minipage}{0.49\textwidth}
			\includegraphics[width=\linewidth]{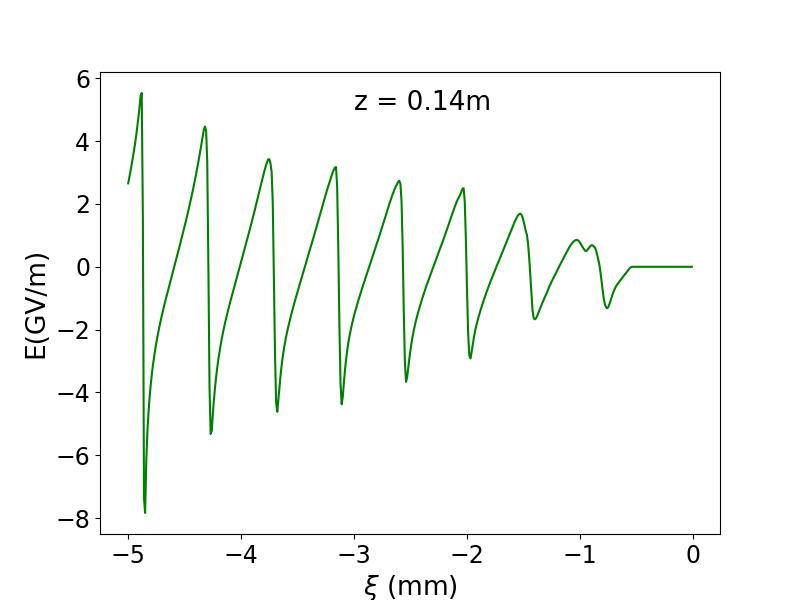}
			\subcaption{}
			\label{fig:Fig_6(d)}
		\end{minipage}
		\caption{(Color) Beam profile (a,b) and the wakefield amplitude (c,d) with co-moving coordinates $\xi$ for \(^{209}\text{Bi}^{83+}\) beam with an RMS beam radius $=$ 1 mm (left) and \(^{209}\text{Bi}^{83+}\) beam with an RMS beam radius $=$ 0.1 mm (right).}
		\label{fig:Fig_6}
	\end{figure*}

	For the Bismuth beam with an RMS radius of 1 mm, self-modulation instability develops in the beam head after 1.4 m, exciting a wakefield with a peak amplitude of 600 MV/m. When the radius is reduced to 0.1 mm, the instability occurs earlier (0.14 m) and drives a significantly stronger wakefield (6 GV/m), as shown in Fig. \ref{fig:Fig_6}..
	
	\subsection{\label{sectionC}Comparison between heavy ion and proton at HIAF}
	
	According to \cite{23}, for very narrow beams, the linear theory expression for the wake amplitude diverges logarithmically with the inverse of the spot size. And this unphysical divergence saturates when the normalized peak beam density $n_b/n_{pe}$ exceeds $ \approx 10 $ for electron drivers and $ \approx 1 $ for positively charged particle drivers. For a three-dimensional Gaussian distributed beam, the maximum beam density appearance where $x = 0, y = 0, z = 0$
	
	\begin{equation}\label{eq:maximum_beam_density}
		n_{max} = n(0, 0, 0) = \frac{N}{ \left(2 \pi \right)^{3/2} \sigma_x \sigma_y \sigma_z}
	\end{equation}
	
\begin{figure}[h]
		\centering
		\begin{subfigure}{0.49\linewidth} 
			\includegraphics[width=1\linewidth]{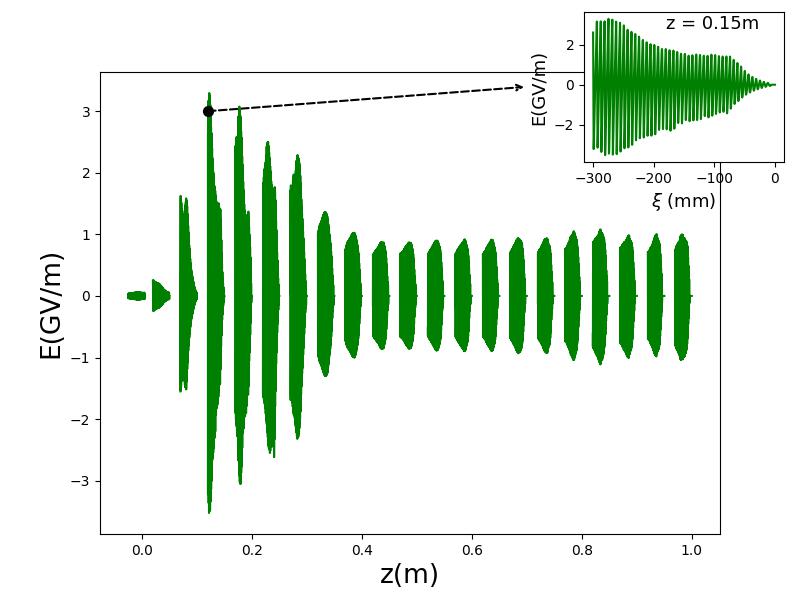}
			\subcaption{}
			\label{fig:Fig_7(a)}
		\end{subfigure}
		\hfill 
		\begin{subfigure}{0.49\linewidth} 
			\includegraphics[width=1\linewidth]{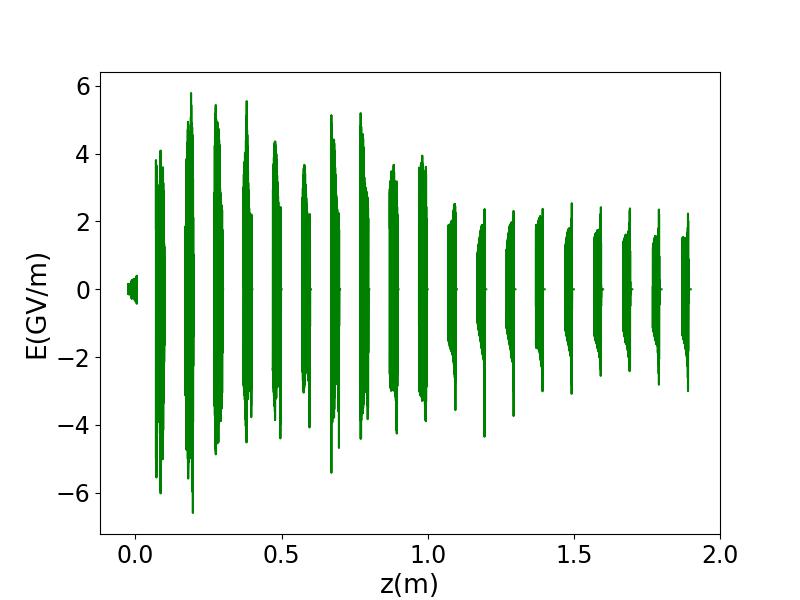}
			\subcaption{}
			\label{fig:Fig_7(b)}
		\end{subfigure}
		\caption{(Color) The evolution of the electric field amplitude at the beam head ($\Delta k_{pe} \xi = 3000$) for the proton (a) / Bismuth (b) beam along the propagation path, captured at different time slices. Each segment corresponds to the electric field amplitude at the beam head for a specific time slice and the horizontal axis represents the position of the beam within the plasma at the corresponding time slice.}
		\label{fig:Fig_7}
	\end{figure}

	Figure \ref{fig:Fig_7} illustrates the evolution of the electric field amplitude at the beam head ($\Delta k_{pe} \xi = 3000$) for the proton (a) / Bismuth (b) beam along the propagation path, captured at different time slices. Each segment corresponds to the electric field amplitude at the beam head for a specific time slice and the horizontal axis represents the position of the beam within the plasma at the corresponding time slice. For the proton beam provided by HIAF, as shown in Table \ref{tab:drivers}, its maximum beam charge density equals its maximum beam density: $ n_{max} = 7.6 \times 10^{13} cm^{-3}$, and the radio of $n_b/n_{pe} \approx 0.03$, leading to a relatively low amplitude wakefield, as shown in Fig. \ref{fig:Fig_7(a)}. For the Bismuth beam with an RMS beam radius $=$ 0.1 mm in HIAF, shown in Table \ref{tab:drivers}, each nucleon carries 83 positive charges, enhancing its ability to attract electrons in the plasma. Therefore, its maximum beam charge density is its maximum beam density multiplied by 83: $83 \times n_{max} = 1.05 \times 10^{14} cm^{-3}$, which is closer to the plasma electron density in Table \ref{tab:drivers}, exciting a high amplitude wakefield, which is shown in Fig. \ref{fig:Fig_7(b)}. Therefore, due to the higher charge density than protons, heavy ion beams can deposit more energy into the plasma and excite wakefields with higher amplitude. 
	Besides that, Fig. \ref{fig:Fig_7(a)} shows that during the self-modulation development, the wakefield amplitude in the head of the proton beam rapidly increases to 3 GV/m and subsequently stabilizes at 1 GV/m. However, the wakefield generated by the heavy ion beam shows different characteristics. The wakefield amplitude in the head of the Bismuth beam first increases to 6 GV/m, then fluctuates slightly over a distance, and finally drops to approximately 3 GV/m, shown in Fig. \ref{fig:Fig_7(b)}. Therefore, the amplitude of wakefield attenuation of Bismuth beam is obviously smaller than that of proton beam. The reason is that the heavy ion beam possesses a heavier particle mass, which results in the plasma wakefield that exerts less adverse disturbance on the Bismuth beam. This increased mass leads to a more stable influence on the wakefield, thereby reducing negative impacts on the Bismuth beam's trajectory and integrity.
	
	\section{\label{section4}Simulations for Electron acceleration}
	
	In this section, the Bismuth beam with an RMS beam radius $=$ 0.1 mm in HIAF will be used for simulating electron acceleration. Considering that the excessive duration of the beam in the simulation leads to a huge demand for time and computational cost, in the following simulations, a series of microbunches (66 microbunches) structure is employed to replace the entire beam to simulate the wakefield generated at the head of the beam. Each Bismuth microbunch has the same peak charge intensity as the numbers shown in Table \ref{tab:drivers} and its bunch length is half of plasma wavelength (for plasma density $n_{pe} = 2.8 \times 10^{15}  cm^{-3}$, corresponding to the plasma wavelength $\lambda_{pe} = 0.628  mm$). The interval between each Bismuth microbunch is also half of the plasma wavelength.
	
	\begin{table}[h]
		\centering
		\caption{\label{tab:HIAF_Bi_acc}
			The parameters of \(^{209}\text{Bi}^{83+}\) and witness beam and plasma for plasma acceleration in HIAF.}
		\begin{tabular}{lc}
			\toprule
			Parameters &HIAF \\
			\hline
			Driver Beam& \(^{209}\text{Bi}^{83+}\)\\
			Energy(GeV/u)& 9.58 \\
			RMS Bunch Radius(mm)& 0.1\\
			Bunch Length(mm)& 0.314\\
			Relative energy spread(\%)& 0.035\\
			Plasma Density($\text{cm}^{-3}$)& $2.8 \times 10^{15}$\\
			Accelerating gradient(GV/m)& 6\\
			\hline
			Witness Beam& electron\\
			Energy(MeV)& 16 \\
			RMS Bunch Radius(mm)& 0.1\\
			Bunch Length(mm)& 0.314\\
			Relative energy spread(\%)& 0.035\\
			\hline
			Simulation grid step& 0.01\\
			Simulation window size(m)& 0.045\\
			particles in layer & 1000\\
			\hline
		\end{tabular}
	\end{table}
	
	For electron acceleration, in order to explore the potential of heavy-ion-driven plasma wakefield acceleration for future collider applications, where simultaneous acceleration of multiple particle bunches is required, three electron bunches are injected with an RMS beam radius of $=$ 0.1 mm and a bunch length equivalent to that of the Bismuth microbunch. The parameters of Bismuth and electron bunch and plasma are summarized in Table \ref{tab:HIAF_Bi_acc}.
	
\begin{figure}
		\centering
		\subfloat[]{\includegraphics[width=8.6cm]{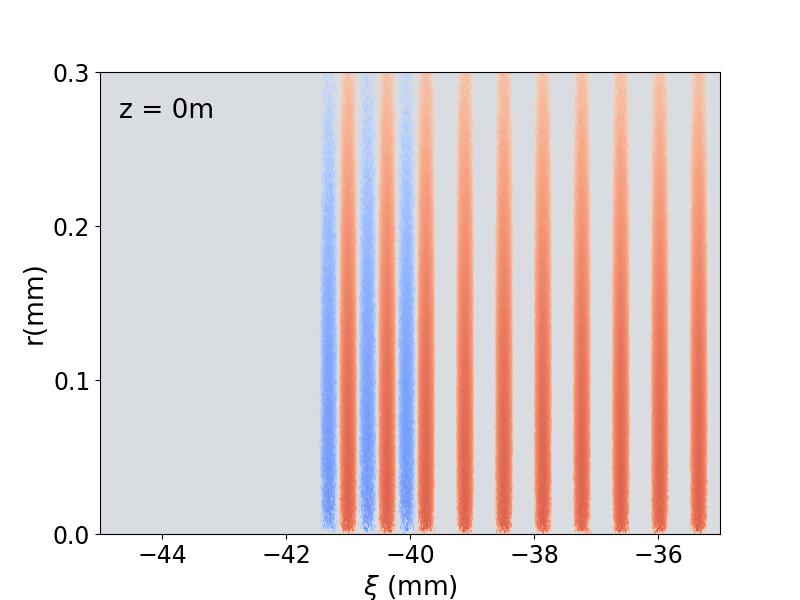}\label{fig:Fig_8(a)}}
		\subfloat[]{\includegraphics[width=8.6cm]{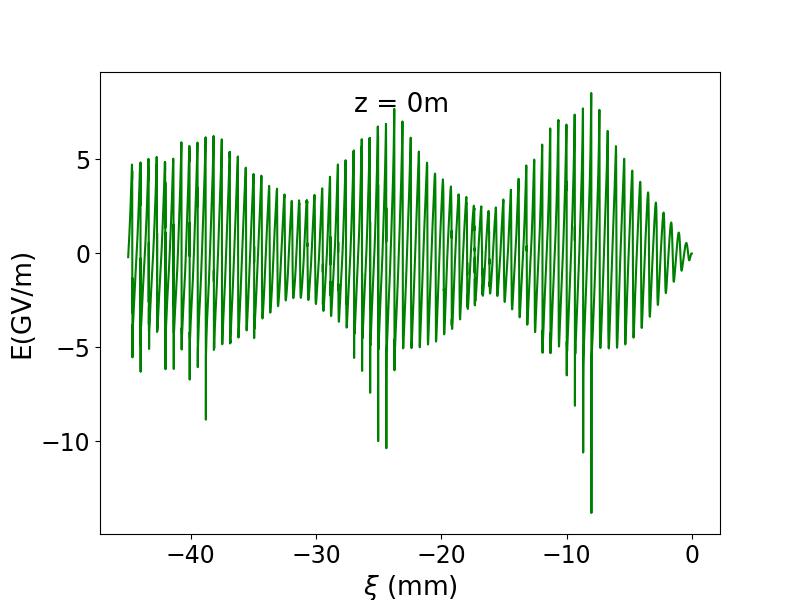}\label{fig:Fig_8(b)}}
		\hfill
		\subfloat[]{\includegraphics[width=8.6cm]{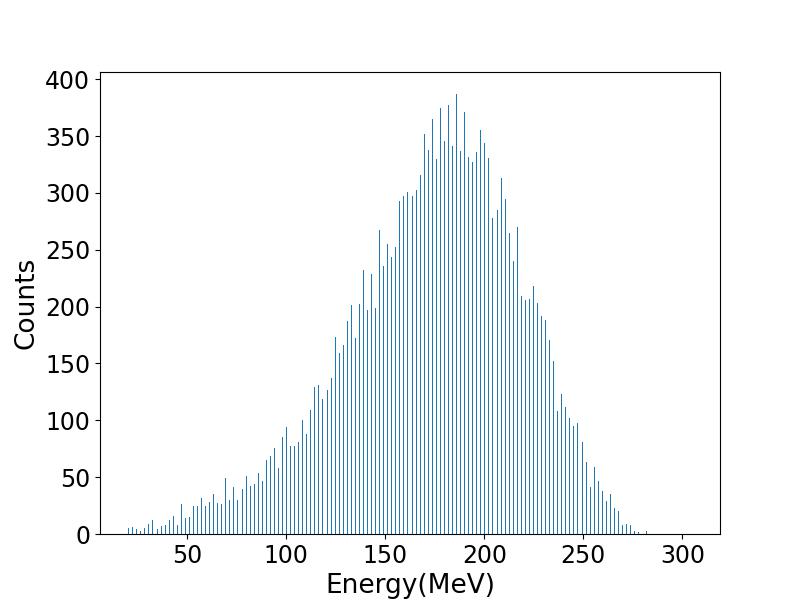}\label{fig:Fig_8(c)}}
		\caption{(Color) Without plasma density gradients, the initial electron (blue) and \(^{209}\text{Bi}^{83+}\) (red) beam distribution (a) , the initial wakefield amplitude in the beam head (b) for \(^{209}\text{Bi}^{83+}\) beam (RMS beam radius $=$ 0.1 mm) with co-moving coordinates $\xi$ and the electron energy distribution after propagating a distance of 0.07 m (c).}
		\label{fig:Fig_8}
	\end{figure}

	Simulation results for electron acceleration are shown in Fig. \ref{fig:Fig_8}. Results show that electron bunches with an energy of 16 MeV can accelerated up to 281.2 MeV after propagating a distance of 0.07 m, and the efficient acceleration gradient is approximately 3.8 GV/m. Due to the velocity of Bismuth beam with an energy of 9.58 GeV/u being relatively lower than electron beam, when electrons are in the process of acceleration, they will experience dephasing process.
	
	\begin{figure}[h]
		\centering
		\includegraphics[width=8.6cm]{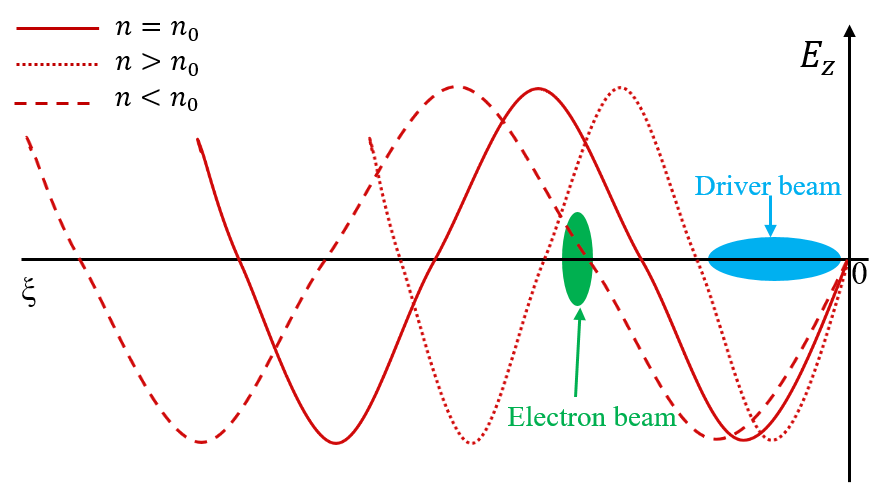}
		\caption{(Color) Phasing of the accelerated electron bunch in plasmas of the proper density (solid line), increased density (dotted line), and reduced density (dashed line).}
		\label{fig:Fig_9}
	\end{figure}

	For fully self-modulated beams in plasma wakefield acceleration, the dephasing length is decided by the relative velocity difference between the driver beam (Bismuth) and witness beam (electron). The dephasing length is 
	
	\begin{equation}\label{eq:dephasing_length}
		L_d = \frac{\lambda_{pe} \beta_w}{4 \Delta \beta}
	\end{equation}
	where $\beta_d$ and $\beta_w$ are the velocities of the driver beam and witness beam, respectively, and $\Delta \beta$ is the relative velocity difference, which can be expressed as = $\Delta \beta = \beta_w - \beta_d$. For the beam parameters in Table \ref{tab:HIAF_Bi_acc}, we can calculate that the dephasing length is about 4.5 cm, which will greatly limit the energy gain of a witness electron beam. After traversing the dephasing length, electrons will exit the focusing and accelerating region, and enter the decelerating or defocusing area, which leads to a degradation in beam quality. 
	
	To investigate a good regime for electron acceleration, plasma density gradients are needed to optimize the simulation \cite{33,34,35}. Plasma nonuniformity can influence the quality of electron acceleration in three aspects: firstly, by altering the dynamics of the heavy ion bunch and the rate of instability growth; secondly, by modifying the trapping conditions; and thirdly, by affecting the properties of the wakefield. Here, we consider the self modulation instability of the Bismuth beam has already fully developed and produced a series of microbunches. Also, the study of trapping conditions for electrons is a complicated field, so we assume the electron bunch is already trapped. Therefore, the main effect of the density gradient comes from dephasing between the wakefield and electron bunches. 
	
	During the development stage of self-modulation instability, the phase of the plasma wave moves backwards for developing instability and the wave can't be used for electron acceleration. After the bunch fully self modulates, the wave can accelerate electrons. In this stage, the phase of the plasma wave responds to density variations by forward or backwards shifts, which is fatal for the witness beam, as illustrated by Fig. \ref{fig:Fig_9}.
	
	In an unperturbed plasma, electrons need to be located somewhere in the region of both focusing and acceleration. If the plasma density increases, the wavelength will shorten, the wave near the electron bunch will shift forward, and the bunch either enters the region of stronger accelerating field or falls into the defocusing region and gets defocused. If the plasma density decreases, the wavelength will elongate, and the bunch may see a lower accelerating field region or even a decelerating field region \cite{36,37}.
	
\begin{figure}[h]
		\centering
		\subfloat[]{\includegraphics[width=8.6cm]{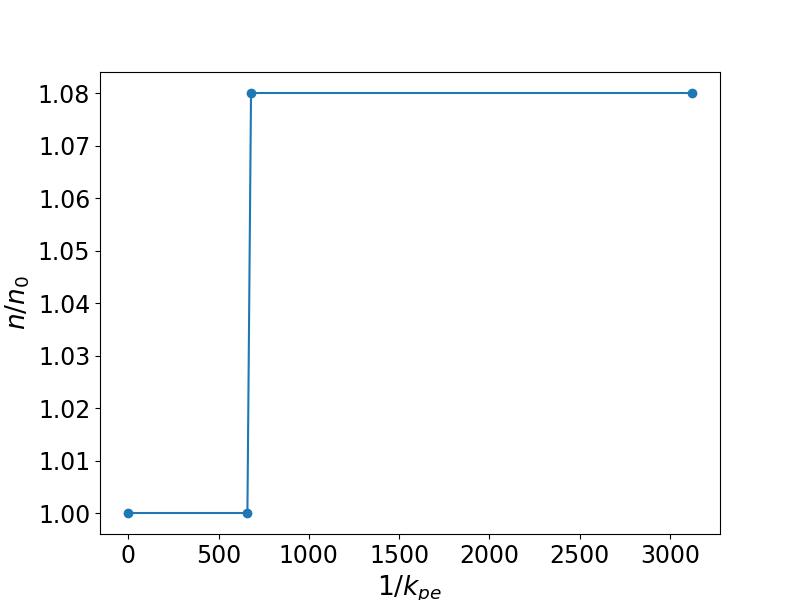}\label{fig:Fig_10(a)}}
		\subfloat[]{\includegraphics[width=8.6cm]{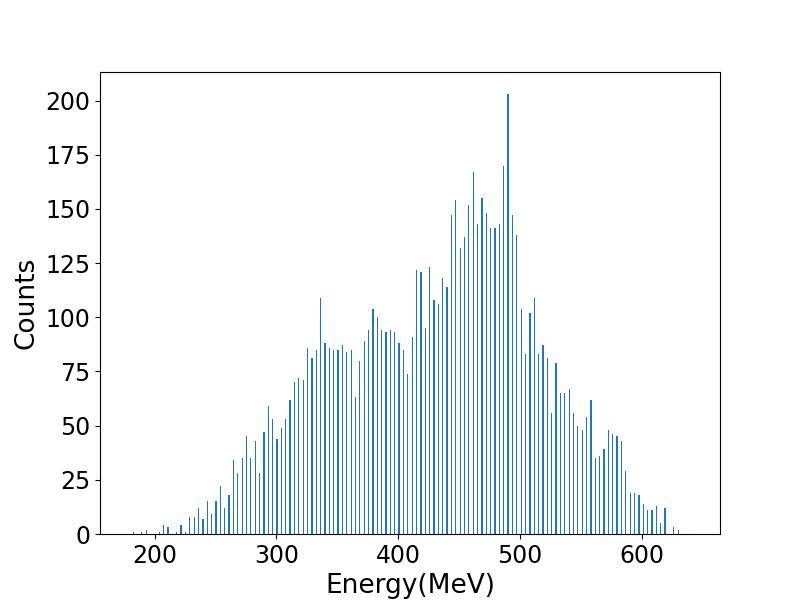}\label{fig:Fig_10(b)}}
		\caption{(Color) Plasma density variation, where $n_0 = 2.8 \times 10^{15} cm^{-3}$(a), the electron energy distribution after propagating a distance of 0.31 m(b)}
		\label{fig:Fig_10}
	\end{figure}

	In order to extend the dephasing length and increase the energy gain of electron beams in plasma wakefield acceleration, plasma density gradients will be introduced. For the variation of plasma density, shown in Fig. \ref{fig:Fig_10}, we set $n_0 = 2.8 \times 10^{15} cm^{-3}$ and increase the plasma density over a distance of 20 $k^{-1}_{pe}$ to form a plasma density step, with $k^{-1}_{pe} = 0.1 mm$ when $n_0 = 2.8 \times 10^{15} cm^{-3}$. Eventually, after propagating a distance of 0.31 m, electron bunches with an energy of 16 MeV can be accelerated up to 636 MeV. However, due to the increasing of plasma density, the phase of the plasma wave excited by each Bismuth microbunches will mismatch and the amplitude of wakefield will weaken, so the effective accelerating gradient is reduced from 3.8 GV/m to 2 GV/m.
	
	\begin{figure}
		\centering
		\subfloat[]{\includegraphics[width=8.6cm]{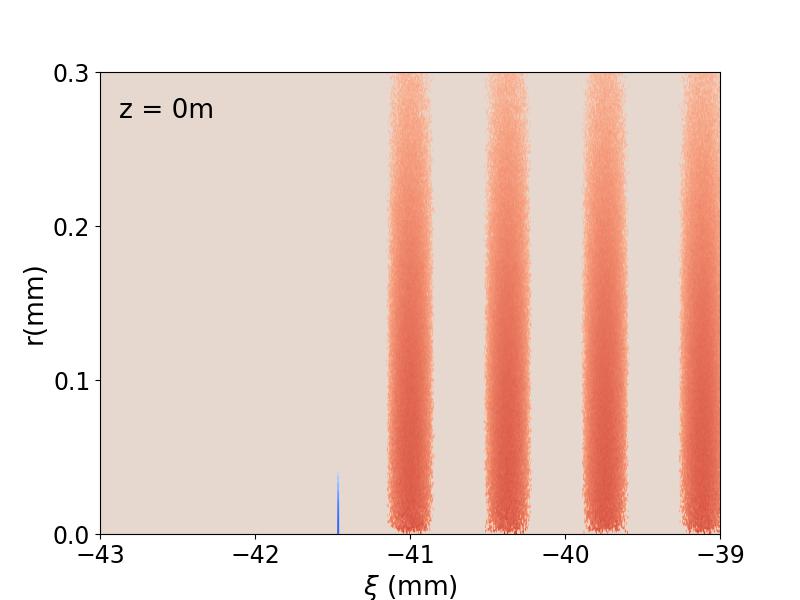}\label{fig:Fig_11(a)}}
		\subfloat[]{\includegraphics[width=8.6cm]{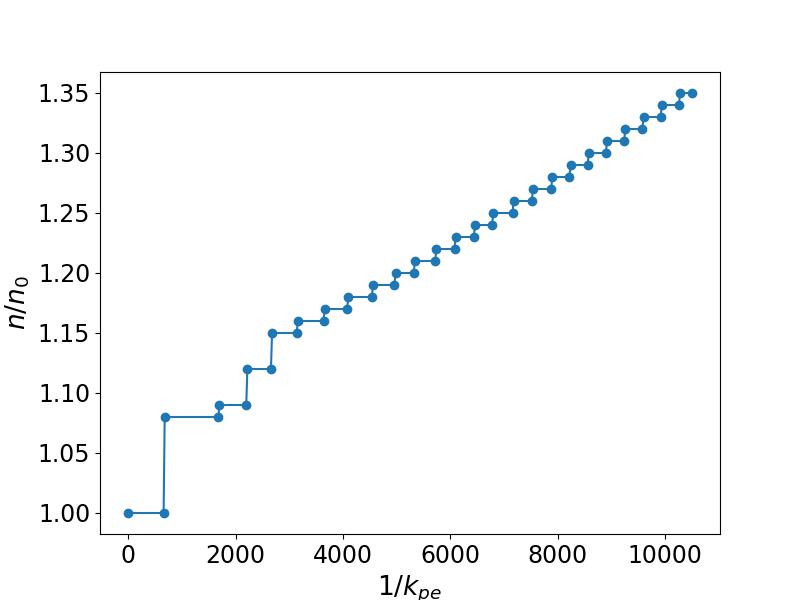}\label{fig:Fig_11(b)}}
		\hfill
		\subfloat[]{\includegraphics[width=8.6cm]{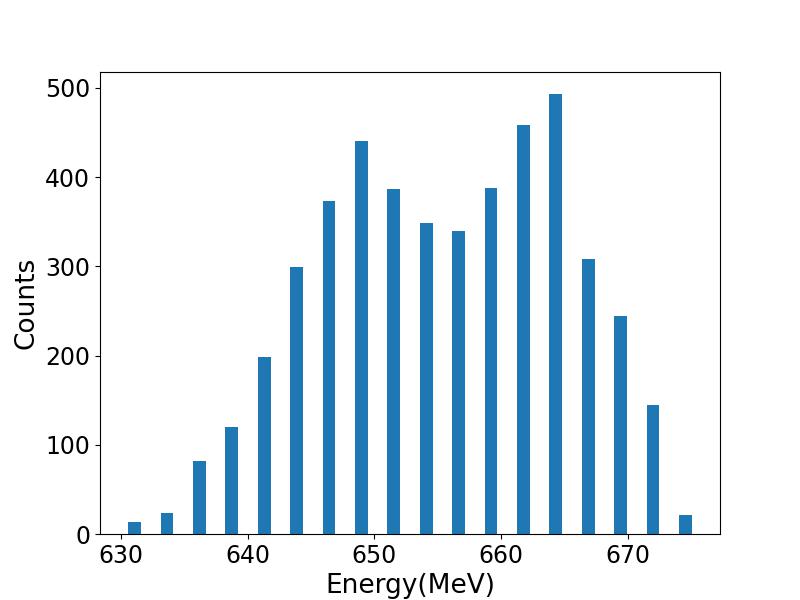}\label{fig:Fig_11(c)}}
		\caption{(Color) With the introduction of plasma density gradients, the initial electron (blue) and \(^{209}\text{Bi}^{83+}\) (red) beam distribution with co-moving coordinates $\xi$ (a), plasma density variation, where $n_0 = 2.8 \times 10^{15} cm^{-3}$ (b) and the electron energy distribution after propagating a distance of 1 m (c).}
		\label{fig:Fig_11}
	\end{figure}
	
	\begin{figure}[h]
		\centering
		\includegraphics[width=8.6cm]{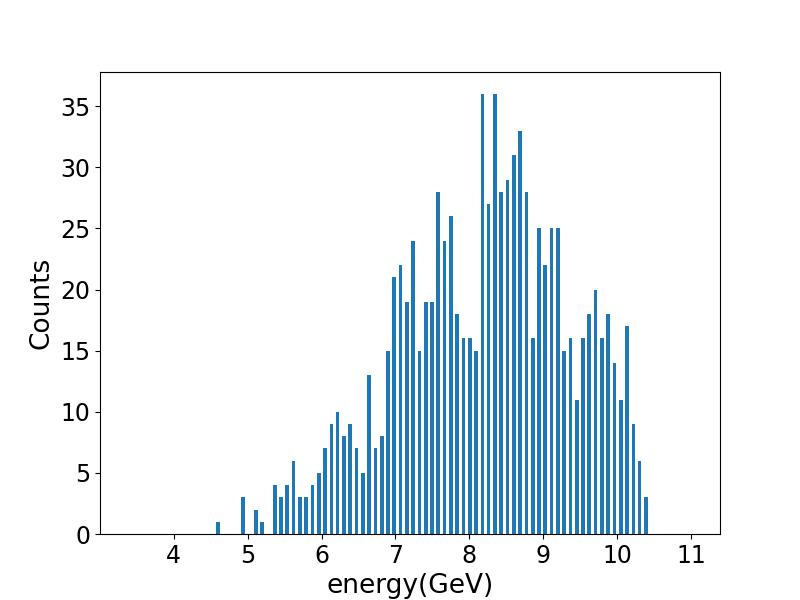}
		\caption{(Color) Without any plasma density gradient, the electron energy distribution after propagating a distance of 2 m.}
		\label{fig:Fig_12}
	\end{figure}
	
	Then, keeping the Bismuth beam and plasma parameters constant, to achieve better beam quality for the electron beam, we consider employing a narrow ($ \approx 10 \mu m$) and short ($ \approx 10 \mu m$) electron bunch for the simulation of electron acceleration. The energy of this electron bunch is 16 MeV and its relative energy spread is 0.035 \%. With some plasma density gradients, after propagating a distance of 1 m in the plasma, electrons can be accelerated from an initial energy of 16 MeV up to 675 MeV, resulting in a good energy spread of 1.5 \%, as shown in Fig. \ref{fig:Fig_11}.
	
	Constrained by the energy of the Bismuth beam, electrons will experience dephasing during the process of acceleration, which may lead to loss due to defocusing, resulting in a degradation in beam quality. Therefore, keeping the electron beam and plasma parameters constant, a series of Bismuth microbunches with an energy in the TeV scale (2.78 TeV) is used to explore the potential of heavy ion driven plasma wakefield acceleration. Fig. \ref{fig:Fig_12} shows that without any plasma density gradient, electron bunches with an energy of 500 MeV can be accelerated up to 10.3 GeV just within a distance of 2 m. By optimizing the witness beam and introducing plasma density gradients, even higher energy gains can be achieved.
	
	\section{\label{section5}Conclusion}
	
	This paper presents the simulation studies investigating a good regime to achieve a high amplitude wakefield for heavy ion driven plasma wakefield acceleration. Among different drivers in HIRFL-CSR and HIAF, the Bismuth beam with an RMS beam radius $=$ 0.1 mm in HIAF, for its high beam charge density and energy, can rapidly develop the self modulation instability in the beam head after a distance of 0.14 m, and excite a wakefield with a maximum amplitude of 6 GV/m. 
	
	Considering the huge demand for time and computational cost, a series of Bismuth microbunches that have the same peak beam charge density and the RMS beam radius instead of the whole Bismuth beam are used to simulate electron acceleration. Limited by the velocity of Bismuth beam in HIAF, electron will experience dephasing, which means electron will exit the accelerating region and see a decelerating or defocusing region. It is crucial to introduce plasma density gradients to ensure that the witness beam will relocate into the accelerating region to extend the dephasing length. With some plasma density gradients, electrons can be accelerated to a maximum of 636 MeV within 0.31 m, as opposed to the maximum of 281.2 MeV achievable after propagating a distance of 0.07 m without plasma density gradients. Then, employing a narrow ($ \approx 10 \mu m$) and short ($ \approx 10 \mu m$) electron bunch for electron acceleration, with some plasma density gradients, after propagating a distance of 1 m in the plasma, electrons can be accelerated from 16 MeV up to 675 MeV, resulting in a good energy spread of 1.5 \%. In the end, a series of Bismuth microbunches with an energy in the TeV scale is also used to explore the potential of heavy ion driven plasma wakefield acceleration. Simulation results show that without any plasma density gradient, electrons with an energy of 500 MeV can be accelerated up to 10.3 GeV just within a distance of 2 m.
	
	In summary, heavy-ion beam drivers hold several potential advantages in plasma wakefield acceleration. The high beam charge density enables to deposit more energy into the plasma and excite a higher amplitude wakefield. Moreover, the heavier mass of the heavy ion particles allows for maintaining a stable wakefield over more extended distances, facilitating the acceleration of the witness beam to even higher energy. However, several challenges and limitations must be addressed. First, the relatively low velocity of heavy ions leads to a short dephasing length, thereby constraining the achievable energy gain during one acceleration stage. Second, we have not considered the impact of emittance on heavy ion-driven plasma wakefield acceleration in this paper. The next step we will employ an external solenoidal magnetic field to achieve plasma wakefield acceleration with large emittance.
	
	\section{\label{section6}Acknowledgments}
	
	The authors would like to appreciate the discussions with K.V.Lotov and the revision comments from Y.J.Yuan. This work is supported by the China National Funds for Distinguished Young Scientists (Grant No. 12425501).
	
	\providecommand{\noopsort}[1]{}\providecommand{\singleletter}[1]{#1}%

\end{document}